\documentclass[sigconf]{acmart}

\usepackage{epsfig}
\usepackage{graphicx}
\usepackage{amsmath}
\usepackage{bbding}
\usepackage{balance}
\AtBeginDocument{%
  }

\setcopyright{acmcopyright}
\copyrightyear{2023} 
\acmYear{2023} 
\setcopyright{acmlicensed}\acmConference[MM '23]{Proceedings of the 31st ACM International Conference on Multimedia}{October 29-November 3, 2023}{Ottawa, ON, Canada}
\acmBooktitle{Proceedings of the 31st ACM International Conference on Multimedia (MM '23), October 29-November 3, 2023, Ottawa, ON, Canada}
\acmPrice{15.00}
\acmDOI{10.1145/3581783.3612530}
\acmISBN{979-8-4007-0108-5/23/10}

\acmSubmissionID{3773}

\begin{document}

\title{High Visual-Fidelity Learned Video Compression}

\author{Meng Li}
\affiliation{%
 \institution{Huawei Technologies}
 \city{Haidian}
 \state{Beijing}
 \country{China}}

\author{Yibo Shi}
\affiliation{%
 \institution{Huawei Technologies}
 \city{Haidian}
 \state{Beijing}
 \country{China}}

\author{Jing Wang \Envelope}
\affiliation{%
 \institution{Huawei Technologies}
 \city{Haidian}
 \state{Beijing}
 \country{China}}
\email{wangjing215@huawei.com}

\author{Yunqi Huang}
\affiliation{%
 \institution{Huawei Technologies}
 \city{Haidian}
 \state{Beijing}
 \country{China}}


\renewcommand{\shortauthors}{Meng Li, Yibo Shi, Jing Wang, \& Yunqi Huang}

%%%%%%%%% ABSTRACT
\begin{abstract}
With the growing demand for video applications, many advanced learned video compression methods have been developed, outperforming traditional methods in terms of objective quality metrics such as PSNR.
Existing methods primarily focus on objective quality but tend to overlook perceptual quality. Directly incorporating perceptual loss into a learned video compression framework is non-trivial and raises several perceptual quality issues that need to be addressed.
In this paper, we investigated these issues in learned video compression and propose a novel High Visual-Fidelity Learned Video Compression framework (HVFVC). 
Specifically, we design a novel confidence-based feature reconstruction method to address the issue of poor reconstruction in newly-emerged regions, which significantly improves the visual quality of the reconstruction. Furthermore, we present a periodic compensation loss to mitigate the checkerboard artifacts related to deconvolution operation and optimization.
Extensive experiments have shown that the proposed HVFVC achieves excellent perceptual quality, outperforming the latest VVC standard with only 50\% required bitrate.

\end{abstract}




\maketitle

\section{Introduction}
The prevalence of smartphones and the accessibility of mobile internet have resulted in the explosive growth of new media short videos. As a result, more and more individuals are producing and sharing video content on various social media platforms.As a result, there is a huge amount of video content uploaded to social media every day, which presents significant challenges for video storage and transmission.Therefore, to further reduce the size of video data while perserving acceptable visual quality has become a crucial research area in video compression.

\begin{figure}[t]
    \centering
    \includegraphics[width=0.5\textwidth]{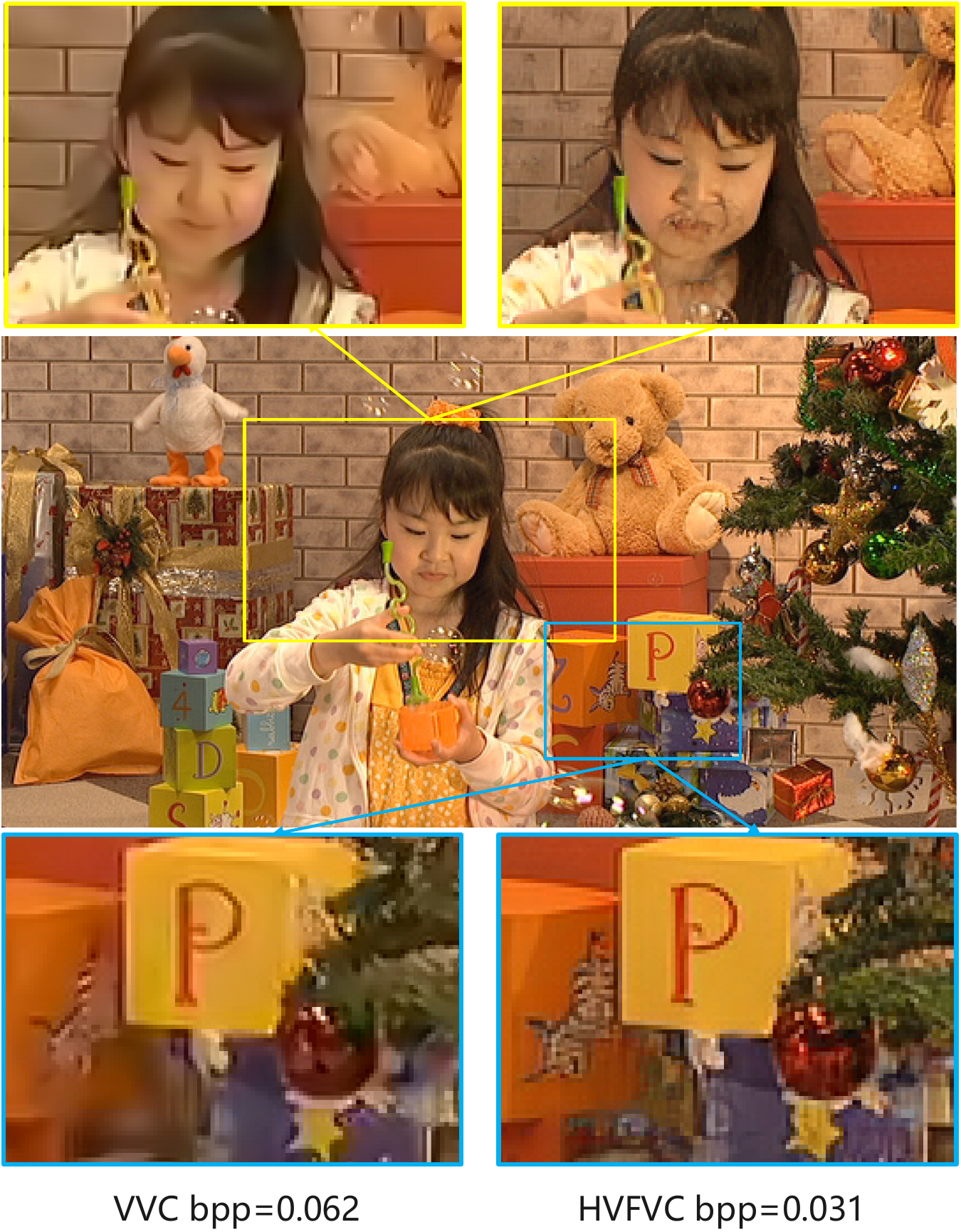}
    \caption{Example of reconstructed frame with HVFVC and VVC(2x bitrate).}
    \label{fig:example}
\end{figure}

% Video compression, which reduces the size of video data while maintaining acceptable visual quality, 
% has become a crucial research area to address these challenges. 

Traditional video coding standards such as AVC, HEVC, and VVC have been proposed to save storage and transmission costs. 
However, since the modules in these standard frameworks including prediction, quantization, and entropy coding, are independently optimized, making it hard to achieve global optimality. In recent years, with the advancement of artificial intelligence (AI) techniques, 
learned video compression methods \cite{DVC,DCN,shi2022alphavc} have emerged as a promising solution to overcome the limitations of traditional video compression methods. 
These methods leverage deep learning algorithms to learn and exploit the inherent patterns and redundancies in video data, 
resulting in higher compression efficiency.
However, most of these approaches only aim to improve the coding efficiency in terms of objective metrics such as PSNR, 
resulting over-smoothing of images as the compression factor increases (lower bitrate). It is believed that a high objective metric score does not necessarily indicate good perceptual quality \cite{mentzer2020high}. Therefore, some studies employ perceptual loss and GAN-based methods to improve the perceptual quality \cite{PLVC,mentzer2022neural, agustsson2022multi}, achieving better visual quality at a lower bitrate. But they still suffer from some criticial issues such as poor quality of newly-emerged regions, checkerboard artifacts, and lower compression ratio. \par
Newly emerged regions in videos refer to areas that become visible or appear in subsequent frames, but were not present in the previous frames. These regions pose a challenge for video compression algorithms as they contain new information that needs to be efficiently encoded while maintaining the overall visual quality of the video. Due to the absence of useful information from reference frames, compressing these regions is more difficult, which makes accurate prediction challenging. Inaccurate prediction usually leads to larger reconstruction errors and wasted bitrates. To this end, we propose a confidence-based feature reconstruction module that identifies and enhances regions of inaccurate prediction, thereby improving the reconstruction quality of newly-emerged regions.
\begin{figure}[t]
    \centering
    \includegraphics[width=0.5\textwidth]{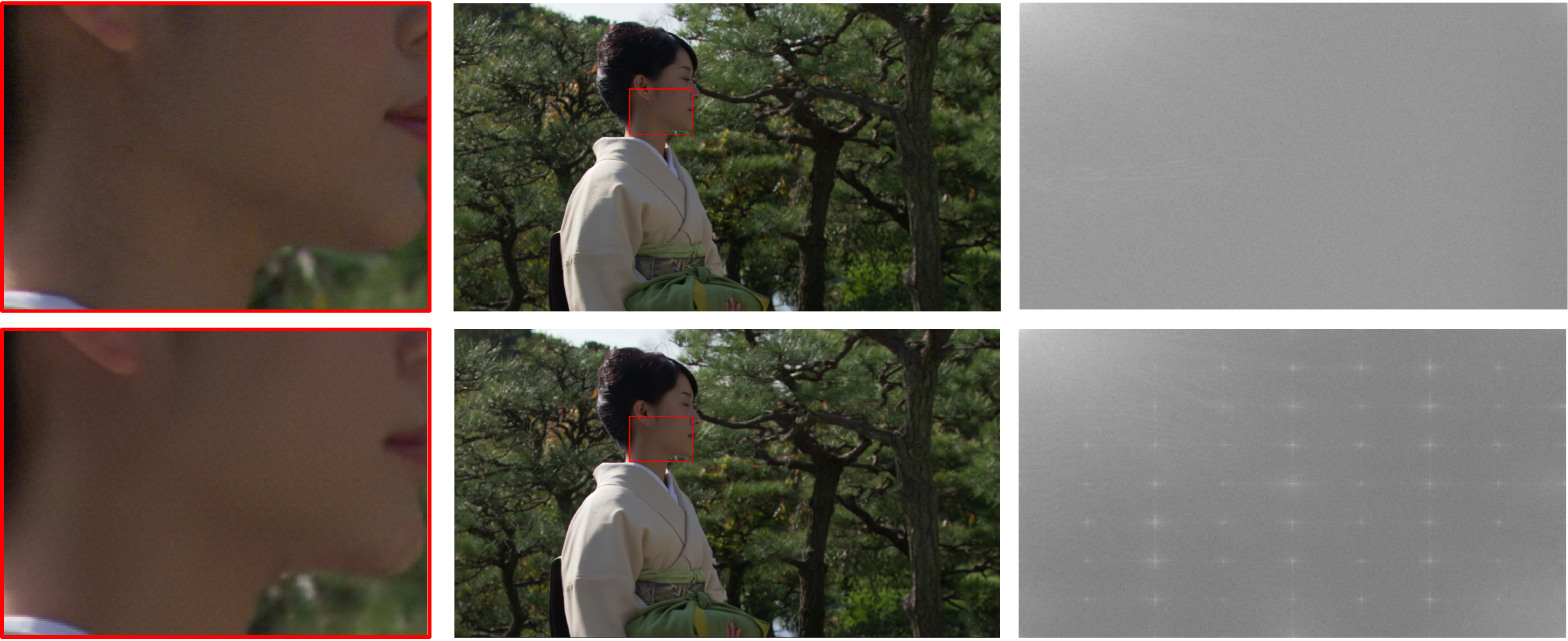}
    \caption{ Comparison between an uncompressed image (top row) and its compressed version with checkerboard pattern (bottom row) in terms of their corresponding spectrum graphs.}
    \label{fig:arti_example}
\end{figure}

The checkerboard artifact is a common problem in low-level tasks \cite{sugawara2019checkerboard}, which is related to the deconvolutional layer \cite{odena2016deconvolution} and the optimization  \cite{sugawara2018super}. It it characterized by regular, checkerboard-like patterns of noise that can severely degrade the visual quality of reconstructed images and videos. The regular noise pattern caused by the checkerboard effect can be clearly observed in the frequency spectrum obtained through DCT transformation, which can be seen from Fig.~\ref{fig:arti_example}. Compared to the spectrum graph of the original frame, the frame containing checkerboard artifacts shows the presence of abnormally bright spots, which are indicative of the periodic texture noise. This problem is particularly challenging in the field of image and video coding, where it can occur at low bitrates. Based on the periodicity of the checkerboard artifact, we propose a periodic compensation loss metric to mitigate such noise.

With these key techniques, we present a novel video compression method, named High Visual-Fidelity Learned Video Compression (HVFVC). Extensive experiments are conducted to evaluate the performance of HVFVC, and the results demonstrate that HVFVC can achieves comparable subjective reconstruction quality with only 50\% bit rate required by the official VVC standard. Fig.~\ref{fig:example} depicts a comparison between the official VVC standard and our proposed algorithm,  in which it is evident that our HVFVC attains much superior visual quality with only half the bitrate.  The contributions of this paper can be summarized as follows:\\
1. We have developed a new framework named High Visual-Fidelity Learned Video Compression, which can reconstruct videos with significantly better visual quality at lower bitrates compared to the current state-of-the-art VVC compression standard. \\
2. We introduce a novel confidence based feature reconstruction to enhance the quality newly emerging regions in videos, while also improving compression efficiency. \\
3. We propose a periodic compensation loss method that effectively eliminates the checkerboard artifact in compressed videos.

% its superior compression performance compared to the official VVC standard at much lower bitrates. 

% In this paper, we present a novel video compression method, named Visual-Friendly Video Compression (VFVC), 
% that addresses the limitations of existing methods and provides a visual-friendly solution for efficient video compression. 
% VFVC incorporates a periodic compensation loss to tackle the checkerboard artifact, 
% an adaptive quantization step strategy to enhance the reconstruction quality of occluded regions, 
% and a high-performance network architecture to further optimize the compression process. 

%The rest of this paper is organized as follows. 
%Section II provides a review of the current research status in video compression, highlighting the limitations of existing methods. 
%Section III presents the proposed HVFVC method in detail, including the periodic compensation loss, the adaptive quantization step strategy, 
%and the high-performance network architecture. Section IV presents the experimental results and performance analysis of VFVC. 
%Finally, Section V concludes the paper and discusses future research directions in video compression.

\begin{figure*}[h]
    \centering
    \includegraphics[width=0.75\textwidth]{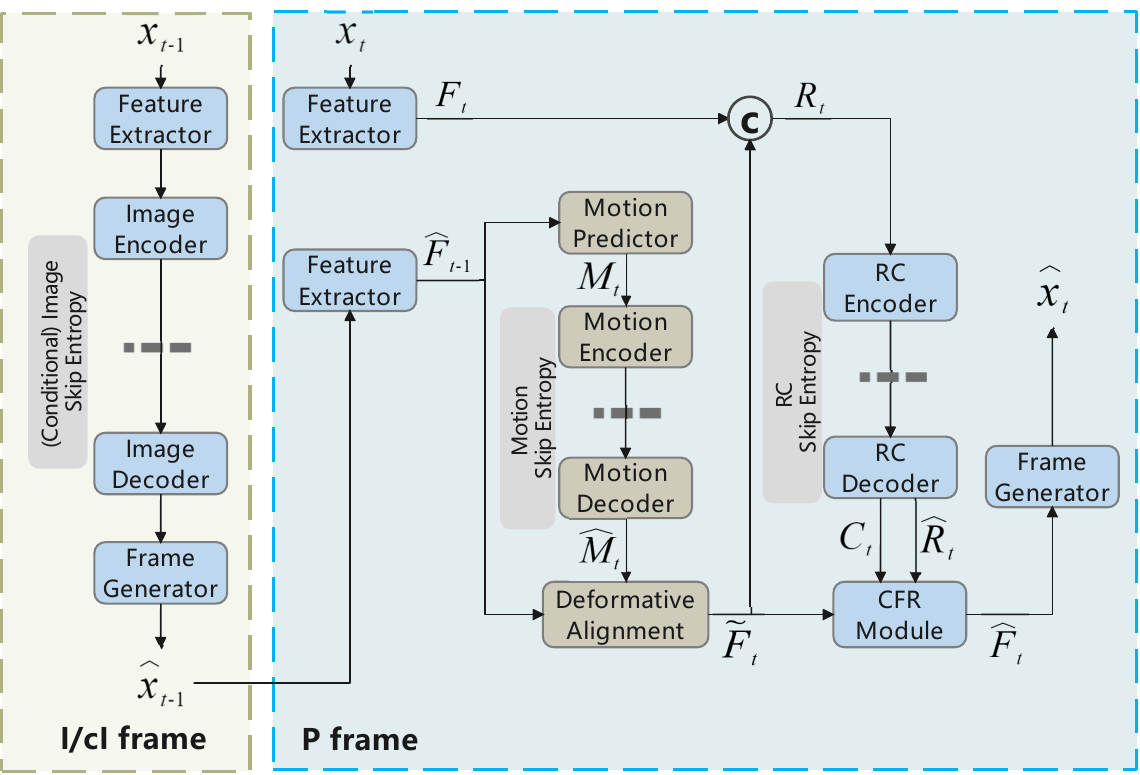}
    \caption{Framework of our method.}
    \label{fig:framework}
\end{figure*}

\section{Related Works}
In the past decades, video compression has been extensively studied and numbers of well-known video compression standards have been proposed \cite{H264,HEVC,VVC}.
Recently, end-to-end learned video compression methods have attracted increasing interest, which show great potential with superior rate-distortion performance.
Compared to traditional methods, learned video compression methods \cite{DVC, DCN, SSF, li2021deep, sheng2022temporal, shi2022alphavc,Yang_2020_CVPR,feng2021versatile} utilize deep learning algorithms to learn and exploit the inherent patterns and redundancies in video data in end-to-end manner, resulting in improved compression performance and visual quality with lower bitrates.

Inspired by traditional video compression framework \cite{H264}, Lu \textit{et. al.} proposed DVC, which compress motion and residual information in an end-to-end optimized manner \cite{DVC}. 
Later, FVC extended this framework into feature space with deformable convolution \cite{DCN}. 
Besides, a series of designs like B-frame \cite{feng2021versatile,pourreza2021extending}, cI-frame \cite{shi2022alphavc}, scale-space flow \cite{SSF}, hierarchical structure \cite{Yang_2020_CVPR} have been proposed to enhance the compression efficiency.

However,minimizing mean squared error (MSE) or other pixel-based distortion metrics, dose not fully align with human perception of image quality. Some recent works have also explored the use of perceptual-guided video compression, where the visual quality of the reconstructed videos is optimized based on human perceptual characteristics. These methods often incorporate perceptual loss functions, such as structural similarity (SSIM) \cite{SSIM} or multi-scale structural similarity (MS-SSIM) \cite{MS-SSIM}, LPIPS \cite{LPIPS}, to guide the compression process and improve the visual quality of the reconstructed videos. 
Besides, GAN-based framework has been proved to be efficent to produce realistic images in the tasks of image compression \cite{PLVC,mentzer2022neural, agustsson2022multi,tian2022coding}.  Mentzer \cite{mentzer2022neural} presents the first neural video compression method based on GAN and to show that it significantly outperforms previous neural and non-neural video compression methods in a user study, showing that the GAN loss is crucial to obtain this high visual quality. Yang \cite{PLVC}  proposes a Perceptual Learned Video Compression (PLVC) approach with recurrent conditional GAN. The authors employ the recurrent autoencoder-based compression network as the generator and propose a recurrent conditional discriminator to generated videos not only spatially photo-realistic but also temporally consistent with the ground truth.  

Although substantial progress have been achieved, there still exists some criticial issues to be addressed, like checkerboard artifacts, lower compression ratio, and over smooth newly-emerged regions.
Our proposed VFVC method builds upon the advancements in learned video compression and introduces novel techniques to address these limitations of existing methods detailed in the following sections. 

\section{Method}
\subsection{Overview}

The architecture of our HVFVC is shown in Fig.~\ref{fig:framework}. 
Inspired by AlphaVC \cite{shi2022alphavc}, our method employs three types of frames: I-frame, P-frame, and cI-frame (conditional I-frame).
The I-frame is solely used for the first frame of each sequence.
For the subsequent GoPs (group of pictures) of the sequence, we begin with cI-frame,
with the aim of reducing the bit-rate and the cumulative error.
Subsequently, we compress the other frames of each GoP using P-frame, 
which significantly reduces temporal redundancy information with respect to I/cI-frame.
The details of the architecture are shown in Fig.~\ref{fig:framework}, and summarized as follows.

$\textbf{I/cI frame}$ 
In our framework, both cI-frames and I-frames use separate encoder and decoder, 
they share the same structure and parameters of encode and decode.
For I-frame, we use the hyper prior as the entropy model \cite{balle2018variational}. 
Differently, the cI-frame uses the reference reconstructed frame as a condition of the entropy
to reduce the inter redundant information, thereby achieving a higher compression rate compared to the I-frame. 
The entropy formulation of I frame and cI frame can be denoted as:

\begin{equation}
R_I\left(\hat{\mathbf{y}}_t \right)=\mathbb{E}_{\hat{\mathbf{y}}_{\mathrm{t}} \sim p_t}\left[-\log _2 q_t\left(\hat{\mathbf{y}}_t\right)\right]
\end{equation}
\begin{equation}
R_{cI}\left(\hat{\mathbf{y}}_t \mid \hat{\mathbf{x}}_{t-1}\right) \\ =  \mathbb{E}_{\hat{\mathbf{y}}_{\mathrm{t}} \sim p_t}\left[-\log _2 q_t\left(\hat{\mathbf{y}}_t \mid H_{\text {align }}\left(\hat{\mathbf{x}}_{t-1}, \hat{\mathbf{M}}_t\right)\right)\right]
\end{equation}
where $R_I$ and $R_{cI}$ denote estimated bitrates of I frame and cI frame, $H_{\text {align }}(\cdot)$ is the method of reconstruction and alignment, $\mathbf{y}_t$ is the quantized latent representation of $\mathbf{x}_t$, and $\hat{\mathbf{M}}_t$ is the motion information. 

\begin{figure*}[h]
    \centering
    \includegraphics[width=\textwidth]{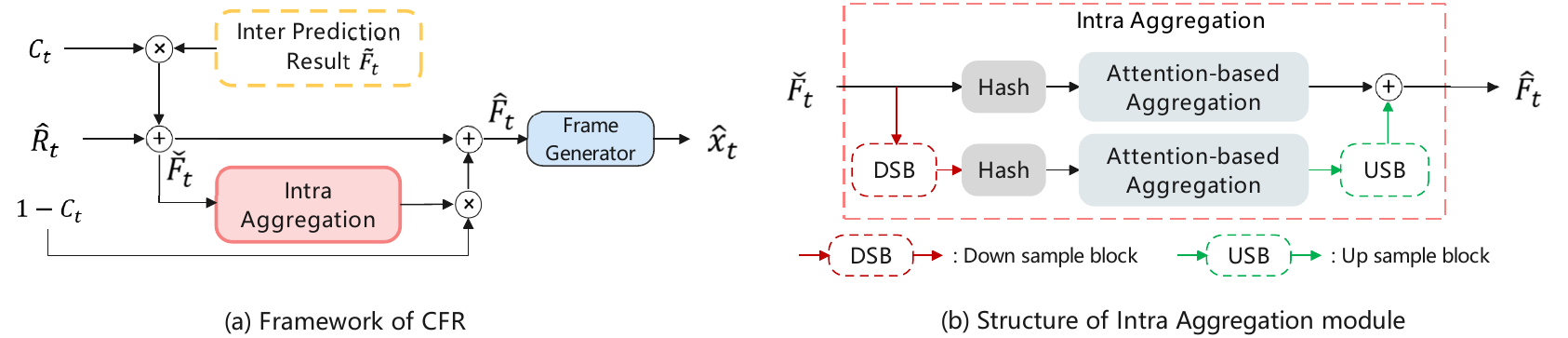}
    \caption{Illustration of the confidence-based feature reconstruction method.}
    \label{fig:RCIA}
\end{figure*}

\begin{figure*}[t]
    \centering
    \includegraphics[width=\textwidth]{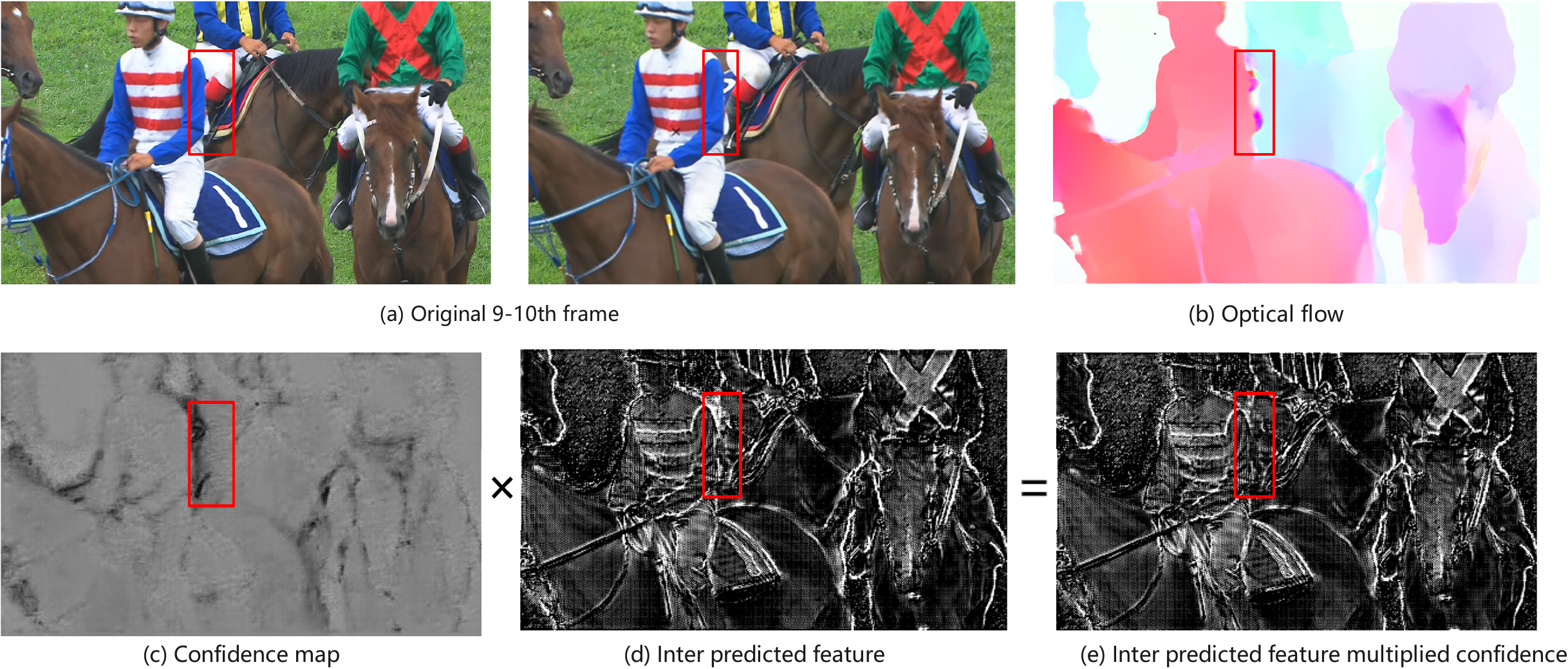}
    \caption{Example of the confidence map and inter prediction. 
        (a): Two adjacent original frames of RaceHorses. 
        (b): Optical flow of these two frames.
        (c,d,e): One channel of confidence map, inter-predicted feature, confidence-based inter-predicted feature.
    }
    \label{fig:confidence}
\end{figure*}

$\textbf{P-frame}$  
For P frame, we use the feature-align based P-frame framework. 
To begin with, the feature extractor module is used to obtain the transformed feature $\mathbf{F}_t$ and $\hat{\mathbf{F}}_{t-1}$ from the current and previous frame, respectively.
Subsequently, the motion predictor module is employed to estimate the motion $\mathbf{M}_t$.
$\mathbf{M}_t$ is then compressed by the motion compression module. 
After that, the predicted feature $\tilde{\mathbf{F}}_t$ is generated by alignment module with 
the reference feature $\hat{\mathbf{F}}_{t-1}$ and the reconstructed motion $\hat{\mathbf{M}}_t$.
Given the predicted feature $\tilde{\mathbf{F}}_t$, the $\mathbf{F}_t$ is compressed to reconstructed feature $\hat{\mathbf{F}}_t$.
Finally, the reconstructed frame $\hat{x}_t$ is reconstructed by $\hat{\mathbf{F}}_t$ through the frame generator. In this framework, the pixel-to-feature motion prediction and probability-based entropy skipping \cite{shi2022alphavc} are used in our method. Furthermore, we have also further optimized the framework to achive better performance, namely confidence-based feature reconstruction, which we will introduce in detail in the following section.

\subsection{Confidence-based Feature Reconstruction}
Given the inter-predicted feature $\tilde{\mathbf{F}}_t$, 
existing learned video compression methods compress and reconstruct $\mathbf{F}_t$ directly through residual information:
\begin{equation}
\begin{aligned}
   & \hat{\mathbf{F}}_t = g_s^r(\mathbf{\hat{r}}_t, \theta_r) + \tilde{\mathbf{F}}_t, \mathbf{\hat{r}}_t = Q(g_a^r(\mathbf{F}_t - \tilde{\mathbf{F}}_t, \theta_r)), \\
   & R(\tilde{\mathbf{F}}_t | \tilde{\mathbf{F}}_t) = \mathbb{E}[-\log_2 p^r(\mathbf{\hat{r}}_t, \theta_r)],
\end{aligned}
\end{equation}
where $g_a^r(,\theta_r), g_s^r(,\theta_r), p^r(,\theta_r)$ is the encoder, decoder and entropy module.
However, due to the unpredictability of certain textures and the limitation of inter prediction methods,
some newly-emerged regions cannot be accurately predicted.
This may result in additional errors and poor reconstructed quality.
To address this issue, we propose a novel feature reconstruction method based on addtional confidence information as shown in Fig.~\ref{fig:RCIA} (a).
The confidence reflects the accuracy of inter-frame prediction, and is used to eliminate inaccurate predictions and enhance the features with poor prediction.
Furthermore, with the confidence, the reconstructed feature is composed of three part: 
confidence-based residual, confidence-based inter-predicted and confidence-based intra-aggregation.
% In the following subsections, we will provide a detailed introduction to these characteristics.
% In addtion, the confidence is extracted and encoded into the bitstream together with the residual.

% the reconstructed feature $\hat{\mathbf{F}}_t$ is generated by compressing the residual:
% For learned video compression, the reconstruction of newly-emerged regions remains a major challenge.
% To this end, we propose a novel feature reconstruction method based on the confidence of residual execution, 
% which consists of two main parts: residual confidence prediction and intra aggregation. 
% The residual confidence indicates the accuracy of inter-frame prediction, 
% and then the features with poor confidence are enhanced using intra aggregation. 
% Next, we will introduce the two parts detailedly.

\subsubsection{Confidence-based residual and inter-prediction} \

% Given the inter-predicted feature $\tilde{\mathbf{F}}_t$, 
% existing learned video codecs typically compress the original feature $\mathbf{F}_t$ directly 
% through compressing the residual:
% \begin{equation}
% \begin{array}{l}
   % \hat{\mathbf{F}}_t = g_s^r(\mathbf{\hat{r}}_t, \theta_r) + \tilde{\mathbf{F}}_t, \mathbf{\hat{r}}_t = Q(g_a^r(\mathbf{F}_t - \tilde{\mathbf{F}}_t, \theta_r)) \\
   % R(\tilde{\mathbf{F}}_t | \tilde{\mathbf{F}}_t) = \mathbb{E}[-\log_2 p^r(\mathbf{r}_t, \theta_r)]
% \end{array}
% \end{equation}
% where $g_a^r(,\theta_r), g_s^r(,\theta_r), p^r(,\theta_r)$ is encoder, decoder and entropy module of the residual compression model.
% However, due to the unpredictability of some textures and the limitation of prediction methods, 
% certain regions can not be accurately predicted, which even may lead to additional errors and poor of reconstructed quality.
% To address this issue, we propose to introduce addtional confidence information that indicates the accuracy of inter-prediction.
The confidence is extracted and compressed into bitstream along with the residual.
With the addtional confidence, the compression process of current feature $\mathbf{F}_t$ can be formulated as followed:
\begin{equation}
\begin{aligned}
    & <\hat{\mathbf{R}}_t, \mathbf{C}_t> = g_s(\mathbf{\hat{w}}_t, \theta_w), \mathbf{\hat{w}}_t = Q(g_a(<\mathbf{F}_t, \tilde{\mathbf{F}}_t>, \theta_w)), \\
    & R(\hat{\mathbf{w}}_t) = \mathbb{E}[-\log_2 p(\mathbf{\hat{w}}_t, \theta_w)], 
\end{aligned}
\end{equation}
where $g_a(, \theta_w), g_s(, \theta_w), p(, \theta_w)$ is the encoder, decoder and entropy module.
The latent representation $\mathbf{w}_t$ is generated from encoder $g_a(, \theta_w)$.
Then it is quantized to $\mathbf{\hat{w}}_t$ and compressed to the bitstream by entropy module and entropy coding.
In the decoder side, the residual information $\hat{\mathbf{R}_t}$ and confidence information $\mathbf{C}_t$ are obtained by the quantized latent representation $\hat{\mathbf{w}}_t$ and decoder $g_s(, \theta_w)$.
After that, the predicted feature is multiplied with the confidence to filter out the inaccurate prediction,
and then added to the residual to get the initial reconstructed feature $\check{\mathbf{F}}_t$:
\begin{equation}
\begin{aligned}
& \check{\mathbf{F}}_t=\tilde{\mathbf{F}}_t \times \mathbf{C}_t+\hat{\mathbf{R}}_t.
\end{aligned}
\end{equation}

As shown in Fig.~\ref{fig:confidence}, 
lower confidence is observed in the regions that cannot be predicted accurately.
This characteristic makes our method to identify and remove the prediction errors, and improve the coding efficency and reconstructed quality of these regions like newly-emerged regions.
In other words, by utilizing the confidence information, our method can degrade to intra coding for these regions.

\subsubsection{Confidence-based intra aggregation} \

Motivated by the intra-prediction technique in traditional coding,
an intra-aggregation module is proposed to improve the efficency of intra-coding.
The goal of this module is to improve the reconstruction of the features with low residual confidence, which can be seen from Fig.~\ref{fig:RCIA}(a). 
The framework can be expressed by:
\begin{equation}
\begin{aligned}
& \hat{\mathbf{F}}_t=\check{\mathbf{F}}_t+f_{ia}\left(\check{\mathbf{F}}_t\right) \times\left(1-\mathbf{C}_t\right)
\end{aligned}
\end{equation}
where $f_{ia}$ denotes the intra aggregation. We selectively apply our intra aggregation method only to regions with low residual confidence, indicating lower accuracy in the prediction. This allows us to enhance the features of these uncertain regions while preserving the high quality of other accurately predicted areas in the video. \par
Our proposed intra aggregation module can aggregate information from global features. 
Specifically,for a given initial reconstrcuted feature $\check{\mathbf{F}}_t=\left\{\mathbf{f}_{t}^i, i \in[1, m\times n] \right\}$, where $m,n$ indicate the width and height of $\check{\mathbf{F}}_t$, we first conduct a search within the entire feature domain, 
selecting $k$  vectors that are nearest to the current feature vector $\mathbf{f}_t^i$. 
Then, we use attention mechanisms to fuse these features together as the enhanced feature. 

Inspired by \cite{kitaev2020reformer,mei2021image, lee2022knn}, we use the spherical Locality Sensitive Hashing for selecting nearest neighbors. 
First a sampled random rotation matrix $\mathbf{R}$ is used to project the vector $\mathbf{f}_t^i$ into a hyper-sphere, and the hash value is defined as: $L(\mathbf{f}_t^i)=argmax([\mathbf{f}_t^i\mathbf{R};-\mathbf{f}_t^i\mathbf{R}])$.
These vectors are evenly distributed into different buckets based on their hash values, 
with vectors that have the same or similar hash values being assigned to the same bucket. 
Finally, the vectors in the same bucket will be enhanced through an attention mechanism, 
for a given vector $\mathbf{f}_t^p$ in a specific bucket $\delta_i$, the output enhanced vector $\mathbf{\hat{f}}_t^p$ can be expressed by:
\begin{equation}
\mathbf{\hat{f}}_t^p  = \frac {\sum_{q \in \delta_i}{\left(s(\mathbf{f}_t^p, \mathbf{f}_t^q) \times g(\mathbf{f}_t^q)\right)}} {\sum_{q \in \delta_i}{s(\mathbf{f}_t^p, \mathbf{f}_t^q)}} \quad \text { s.t. } \quad\left\|\delta_i\right\|_0 \leq k
\end{equation}
where $s(\mathbf{f}_t^p,\mathbf{f}_t^q)$ computes the similarity between the input $\mathbf{f}_t^p$ and $\mathbf{f}_t^q$, $g()$ means feature transformation, and $k$ is the bucket size.   Additionally, to locate analogous features at different scales and facilitate feature fusion, we employed a dual-scale intra-frame fusion method, as depicted in Fig.~\ref{fig:RCIA}(b).

\begin{figure}[t]
    \centering
    \includegraphics[width=0.48\textwidth]{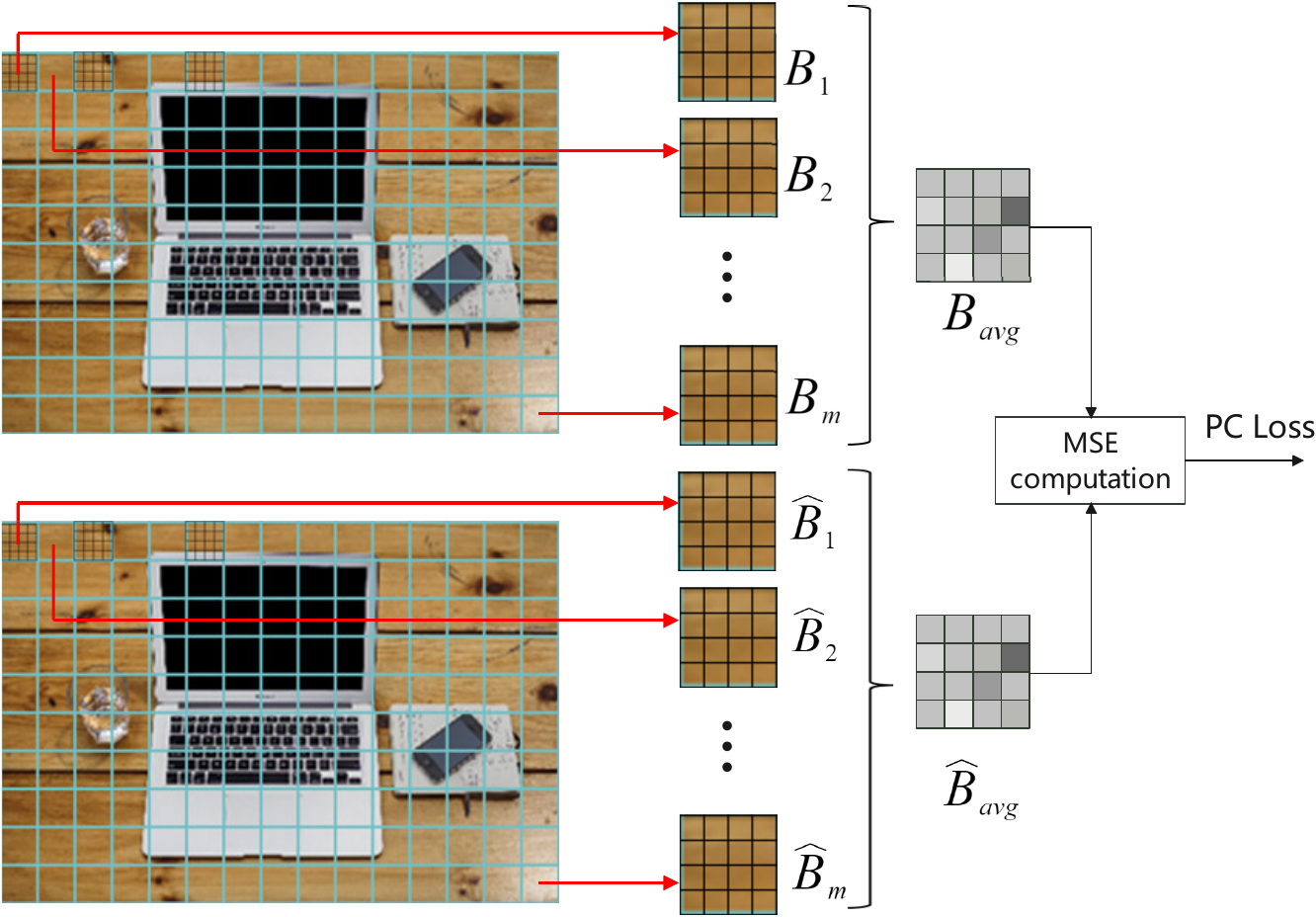}
    \caption{Flowchart of the computation of the periodic compensation loss.}
    \label{fig:pc}
\end{figure}

\begin{figure*}[t]
    \centering
    \includegraphics[width=0.97\textwidth]{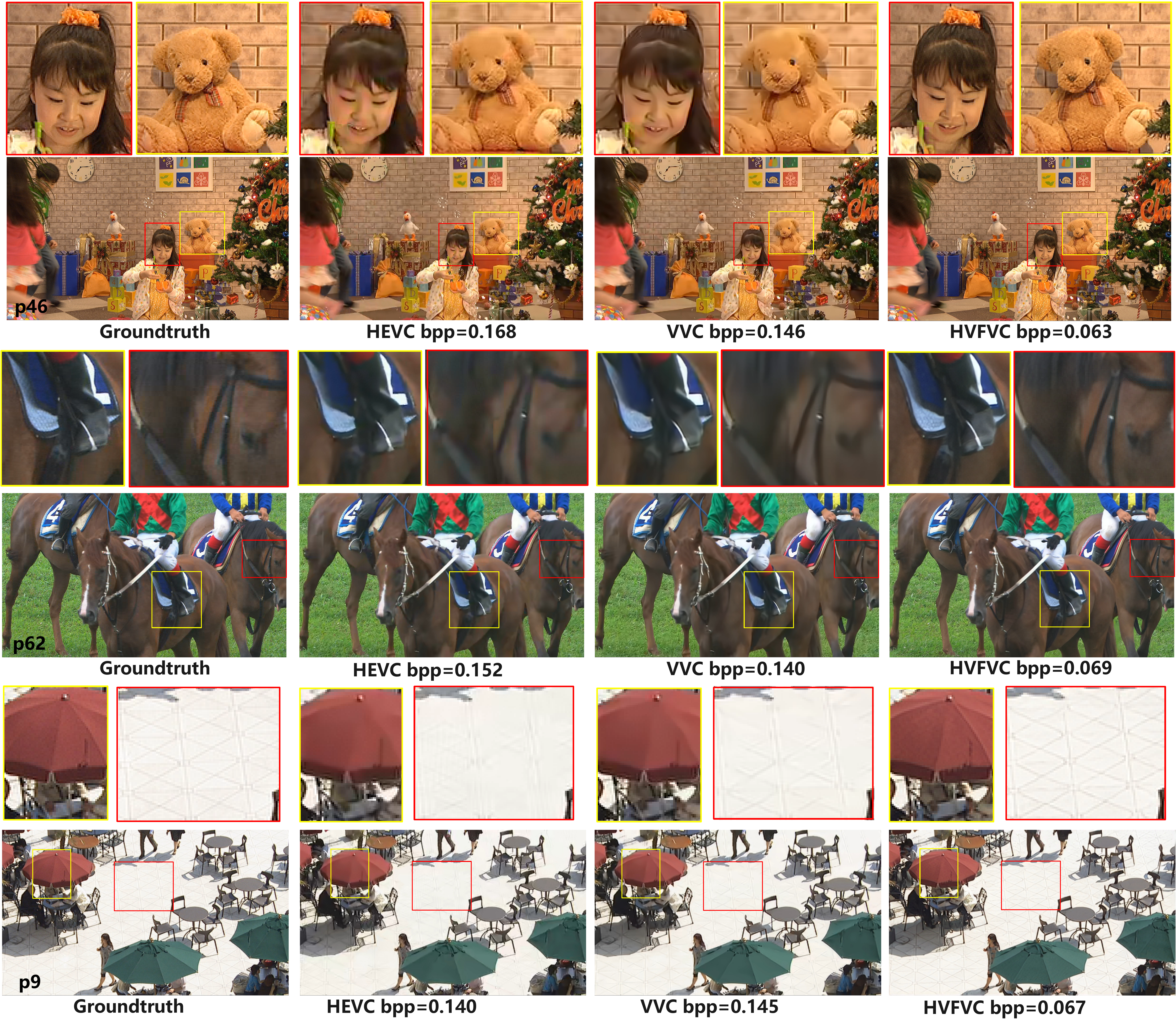}
    \caption{Comparison of different methods.}
    \label{fig:cmp}
\end{figure*}

\subsection{Loss Function}
For each trained frame including I/cI frame and P frame, we use the mixed perceptual loss including L1 loss, LPIPS loss and GAN loss. LPIPS is a loss metric that makes images more semantically similar, which is more consistent with human perception. And GAN is widely used for perceptual optimization.
Therefore, adversarial training is adopted to produce visual friendly frames. Specifically, we employ the relativistic average discriminator, which has been proven to be effective in producing photo-realistic image \cite{li2022content}. The generator loss $\mathcal{L}_{D}^{R a}$ and the discriminator loss $\mathcal{L}_{G}^{R a}$ are formulated as:
\begin{equation}
\label{Ragan}
\begin{aligned}
    \mathcal{L}_{D}^{R a}\left(x, \tilde{x}\right)=-\mathbb{E}_{x}\left[\log \left(D_{R a}\left(x, \tilde{x}\right)\right)\right]-\mathbb{E}_{\tilde{x}}\left[\log \left(1-D_{R a}\left(\tilde{x}, x\right)\right)\right]\\
    \mathcal{L}_{G}^{R a}\left(x, \tilde{x}\right)=-\mathbb{E}_{x}\left[\log \left(1-D_{R a}\left( x,\tilde{x}\right)) \right]-\mathbb{E}_{\tilde{x}}\left[\log \left(D_{R a}\left(\tilde{x},x \right))\right]\right.\right.
\end{aligned}
\end{equation}
where $D_{R a}\left(x, \tilde{x}\right)=\sigma\left(C\left(x\right)-\mathbb{E}_{\tilde{x}}\left[C\left(\tilde{x}\right)\right]\right)$. $C\left(x\right)$ and $C\left(\tilde{x}\right)$ are the the non-transformed discriminator ouput. \par
However, utilizing this mixed perceptual loss function increases the likelihood of generating checkerboard artifacts in the reconstructed frame \cite{sugawara2018super}, particularly at low bitrates.  Therefore, we use the proposed periodic compensation loss metric to combat the checkerboard artifact, which will be introduced detailedly.\par
$\textbf{Periodic compensation loss}$ Checkerboard artifacts are characterized by periodic, checkerboard-like patterns of noise, which severely degrade the visual quality of the reconstructed frames. A reasonable way to eliminate them is to add an extra constraint based on their period. The schematic diagram of this method is shown in Fig.~\ref{fig:pc}, which illustrates the steps involved in constructing the proposed loss. First, the original image and the reconstructed image are divided into small blocks $\mathbf{B}_i$ and $\mathbf{\hat{B}}_i, i \in [1,m]$,  based on the period of the checkerboard artifact. Then, the mean value of pixels at corresponding positions in all small blocks is calculated, resulting in the final aggregated feature blocks: $\mathbf{B}_{a v g}=\frac{1}{m} \sum \mathbf{B}_i$, $\mathbf{\hat{B}}_{a v g}=\frac{1}{m} \sum \mathbf{\hat{B}}_i$. Since each element in the feature blocks is obtained by uniformly sampling the entire image, it can be approximated as independent of the image content. This approach partitions the original and reconstructed images into blocks based on the periodicity of the checkerboard artifact noise, thereby effectively isolating the influence of image content and providing a more targeted constraint on the checkerboard noise. Next, we compute the mean squared error (MSE) between the feature blocks of the original frame and the reconstructed frame, which serves as our periodic compensation loss (PC loss): $\mathcal{L}_{PC} =\mathbb{E}\|\mathbf{B}_{a v g}-\mathbf{\hat{B}}_{a v g}\|_{2}^2 $.
\par
The final loss function is targeted to optimize the rate-distortion-perception trade-off. The whole loss is formulated as:
\begin{equation}
    \mathcal{L} = \lambda R + \alpha \mathcal{L}_{l1}(x,\hat{x}) + \beta \mathcal{L}_{lpips}(x,\hat{x}) + \gamma \mathcal{L}_{GAN}^{R a}(x,\hat{x}) + \phi \mathcal{L}_{PC}(x,\hat{x})
\end{equation}
where $ R$, $\mathcal{L}_{l1}$, $\mathcal{L}_{lpips}$, $\mathcal{L}_{GAN}^{R a}$ denote the estimated bitrate, the L1 loss, the LPIPS loss and the GAN loss,  and $\lambda$, $\alpha$, $\beta$, $\gamma$, $\phi$ are hyper parameters. And we use the feature maps from the VGG16 \cite{simonyan2014very} network  pre-trained on ImageNet \cite{deng2009imagenet} to calculate LPIPS.

\begin{figure*}[t]
    \centering
    \includegraphics[width=\textwidth]{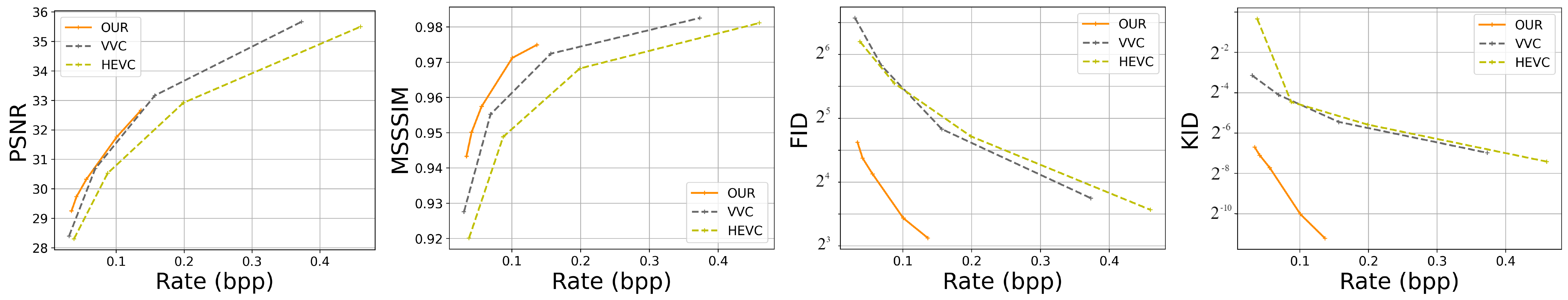}
    \caption{Comparison of rate-distortion and -perception curves}
    \label{fig:metrics}
\end{figure*}

\section{Experiment}
\subsection{Training details}
The datasets used for training in our method is the Vimeo-90k dataset \cite{xue2019video} consisting of 4278 videos with 89800 independent shots. It is notoriously difficult to train GANs for frame generation, so the training procedure is divided into two stages. In the first stage, we optimized the video compression model for 60 epoches using th
e MSE loss metric, with PSNR as the primary performance metric. With the pre-trained model of the first stage, we train the perceptual optimized model for 10 epoches with the loss function Eq.(9) mentioned in section 3.2 in the second stage.  During the training phase, we randomly crop the images into patches of size 256 × 256 and set the batch size to 8. In the first stage, the learning rate was initially set to $1e-4$ and reduced to $5e-5$ in the 31st epoch. And in the second stage, the learning rate is set to $1e-5$. The hyper parameters are set as: $\alpha=1e-2, \beta=1, \gamma=5e-4, \phi=0.1$, $\lambda$ is changing during the training to get the target bitrate.  And We evaluate our proposed algorithm on the HEVC datasets \cite{bossen2013common}, which contains videos with different resolution 416 × 240, 832 × 480 and 1920 × 1080. The GOP size is set to 50 for all testing datasets.\par

In order to get the target bitrates of the videos, we propose a rate control strategy. We set a target bitrate and a validation video, and during training, we tested the bitrate of the validation video every 1000 iterations. If the actual bitrate was greater than 110\% of the target bitrate, we increased $\lambda$ by 1.1, and if it was less than 90\%, we decreased $\lambda$ by 0.9. Employing this strategy allowed us to effectively obtain a model that met a specific target bitrate.

\subsection{Visual Result}
In this section, we showcase the advantages of our algorithm by comparing it to HEVC \cite{HEVC} and the latest VVC \cite{VVC}. For HEVC and VVC, we use the IPP mode that only refers to the previous frame when encoding a new frame.   The results of the comparison are depicted in Fig.~\ref{fig:cmp}, in which we demonstrate the reconstruction results of the three algorithms on various objects, including faces, toys, animals, and still objects. To demonstrate the superiority of our algorithm, we set the bitrates of VTM and HM to be more than twice the bitrate of our HVFVC. It is obvious that our algorithm's video reconstruction quality surpasses that of HEVC and VVC, even with only half the bitrates. For example, both HEVC and VVC reconstructions suffer from blurriness, leading to a loss of texture details (e.g. the face, the toy, the floor). In contrast, our HVFVC can restore a greater number of texture details, resulting in an improved overall reconstruction visual quality. Moreover, HEVC and VVC reconstructions exhibit color bias in the reconstruction of some areas(e,g, the face, the horse), whereas our algorithm produces reconstructions that are more close to the original video. Overall, HVFVC demonstrates superior performance in terms of color fidelity and texture details, resulting in more realistic and visually appealing reconstructions.

\subsection{Statistic Result}
In order to quantify our performance, we adopt the widely used metrics such as PSNR, MSSIM, Frechet Inception Distance (FID)\cite{FID} and Kernel Inception Distance (KID) to evaluate our method. While PSNR and MSSSIM measure the similarity between individual image pairs, FID and KID assesses the similarity between the distribution of reference images and the generated or distorted images. FID and KID are widely used metrics for evaluating the subjective results of super-resolution or generative tasks. The R-D curves of HVFVC, AlphaVC, PLVC, HEVC and VVC are shown in Fig.~\ref{fig:metrics}. Our HVFVC exhibits superior performance over HEVC and VVC in terms of metrics including MSSSIM, FID and KID, indicating its effectiveness in achieving high visual quality. Moreover, our algorithm also achieves comparable PSNR performance to VVC while surpassing HEVC in this regard. These results demonstrate the potential of our algorithm in enhancing video coding and compression applications.

\begin{figure}[t]
    \centering
    \includegraphics[width=0.5\textwidth]{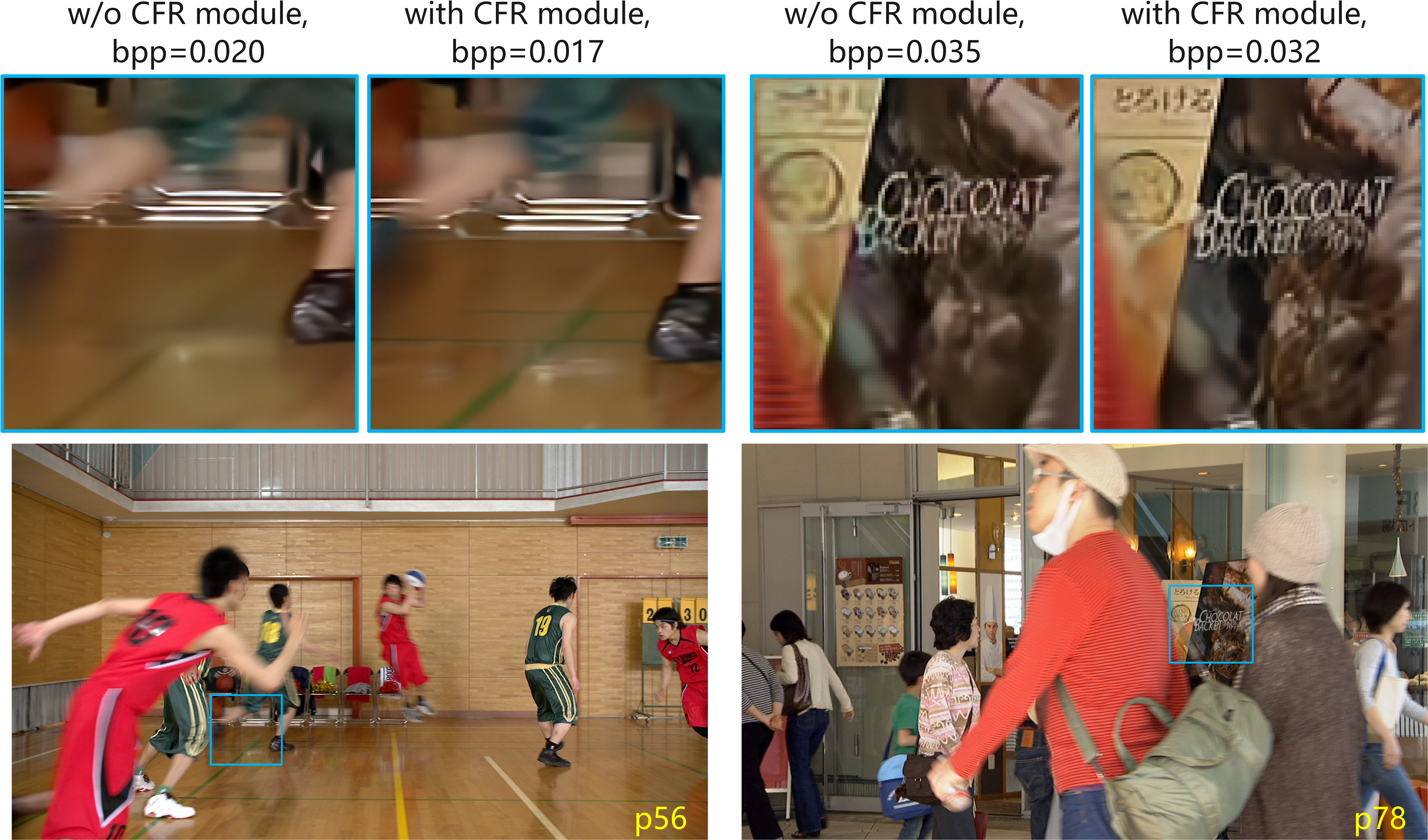}
    \caption{Ablation on the CFR module.}
    \label{fig:rcia_ablation}
\end{figure}

\begin{figure*}[t]
    \centering
    \includegraphics[width=0.95\textwidth]{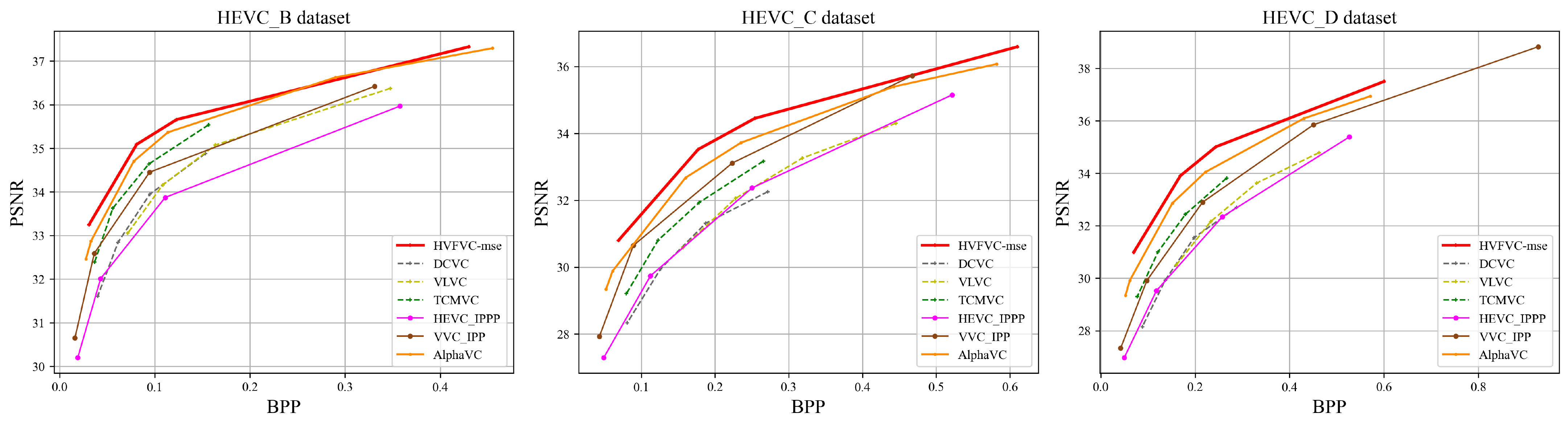}
    \caption{Comparison of RD curves in terms of PSNR}
    \label{fig:rd}
\end{figure*}

\subsection{Ablations}
In this section, we show the effectiveness of the confidence based feature reconstruction (CFR) and the periodic compensation loss. To this end, we first compared the visual differences between the reconstructed videos with and without the CFR module, as shown in Fig.8. As can be seen in the Fig.~\ref{fig:rcia_ablation} that with CFR, the model can reconstruct more details, expecially in the newly emerged regions.  In the first scenario, the boundary lines on the basketball court floor is initially occluded in the previous few frames. However, with the use of the CFR module, our method can accurately reconstruct the texture, resulting in a clear and distinct appearance. In contrast, the non-CFR model exhibits inferior performance in reconstructing this texture. In the second scenario, a similar observation can be made for the advertisement billboard, which is initially occluded in the first few frames. By incorporating the CFR module, our algorithm is able to reconstruct the text on the billboard with greater clarity and detail, compared to non-CFR models. These results demonstrate the effectiveness of our algorithm in handling newly-emerged regions and recovering fine texture details. \par

\begin{figure}[t]
    \centering
    \includegraphics[width=0.5\textwidth]{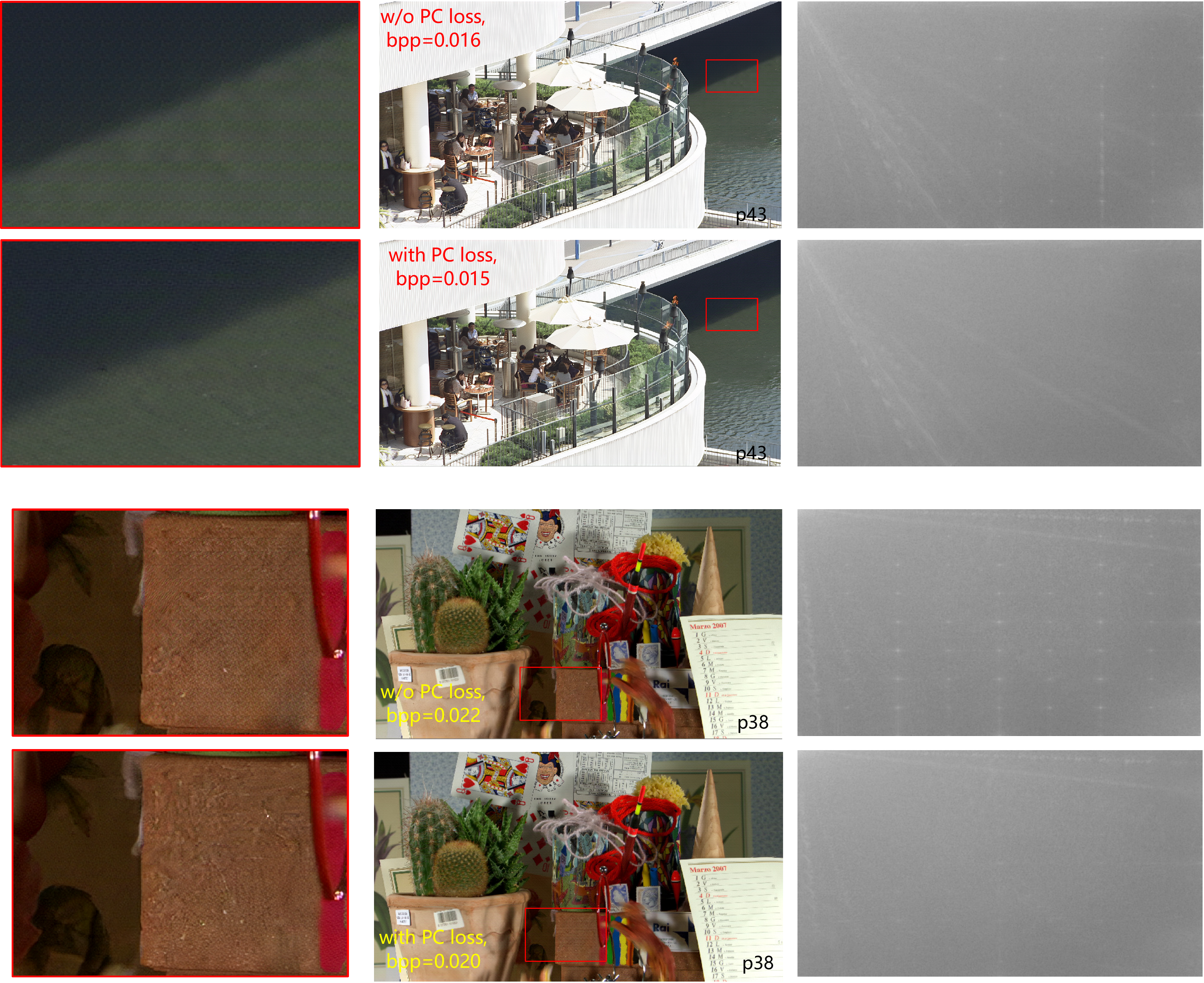}
    \caption{Ablation on the period compensation loss metric.}
    \label{fig:pc_ablation}
\end{figure}

\begin{figure}[t]
    \centering
    \includegraphics[width=0.43\textwidth]{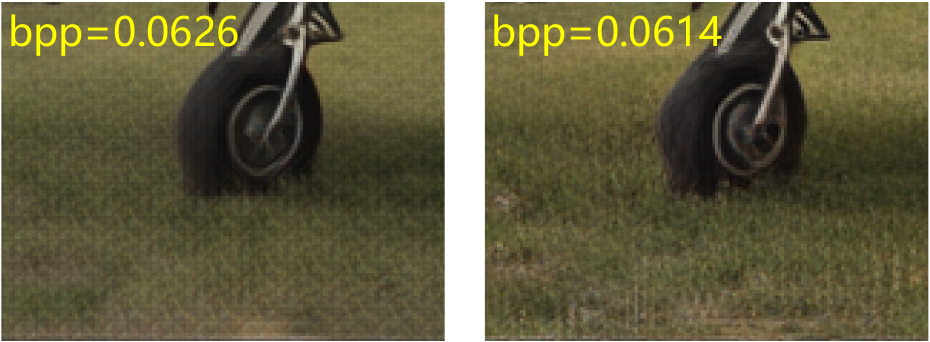}
    \caption{Example of the effectiveness of the periodic compensation loss on image compression.}
    \label{fig:kodim20_arti1}
\end{figure}

Besides, to further demonstrate the effectiveness of our CFR module, we compared the MSE-optimized result of our HVFVC (the first training stage mentioned in Section 4.1) with the non-CFR version (or AlphaVC). The rate-distortion curves in terms of PSNR for Class B, Class C, and Class D of the HEVC dataset \cite{bossen2013common} are shown in Figure~\ref{fig:rd}. With the help of CFR module, we can save about 17.09\% BDRate. This is because the CFR module can make full use of global spatial information to reconstruct more accurate features, especially for newly emerged regions. This module can serve as a good supplement to inter-prediction, thus improve the compression performance. We have also shown several other methods for reference, including DCVC \cite{li2021deep}, VLVC \cite{feng2021versatile}, TCMVC \cite{sheng2022temporal}, HEVC \cite{HEVC}, VVC \cite{VVC}, and AlphaVC \cite{shi2022alphavc}.  It is clear that our result outperforms all the others, thus proving the effectiveness of our method.

In addition, we also demonstrated the effectiveness of the periodic compensation loss in our approach. Specifically, we trained two models with our architecture, differing only in the use of PC loss. The visual difference between the reconstructed videos with and without the PC loss is shown in Fig.~\ref{fig:pc_ablation}. In the top rows of the two examples, we observed significant checkerboard artifacts, resulting in bad visual quality in the frame and abnormally bright spots in its frequency spectrum graph. However, with the help of PC loss, the local textures (on the bottom parts) appear more natural and the obnormal bright spots on the frequency spectrum graphs disappear. Furthermore, our PC loss can also be applied to various low-level tasks, such as image/video compression, super-resolution, and restoration, to mitigate the common issue of checkerboard artifacts. Specifically, the effectiveness of the periodic compensation loss in image encoding is demonstrated in Fig.~\ref{fig:kodim20_arti1}, where the compression model follows the approach proposed in reference \cite{li2022content}. Overall, our proposed periodic compensation loss can effectively address the problem of checkerboard artifacts.

\section{Conclusion}
In this paper, we propose a novel High Visual-Fidelity Learned Video Compression method, in which we present a new module named confidence-based feature reconstruction. This module can not only improve the quality of the reconstructed newly-emerged regions, but also improves the coding efficiency. Besides, to combat the well-know checkerboard artifact, we propose the periodic compensation loss. Experiments clearly show the superiority of our method on visual quality as well as different metrics. 

\bibliographystyle{ACM-Reference-Format}
\balance
\bibliography{sample-base}

\appendix
\section{Additional results}

%Due to the constraints imposed by the length of the paper, we have included supplementary materials containing additional experiments for the purpose of providing readers with further references.

\subsection{Comparion of different optimizition approaches}
\begin{figure}[ht]
	\centering
	\includegraphics[width=0.5\textwidth]{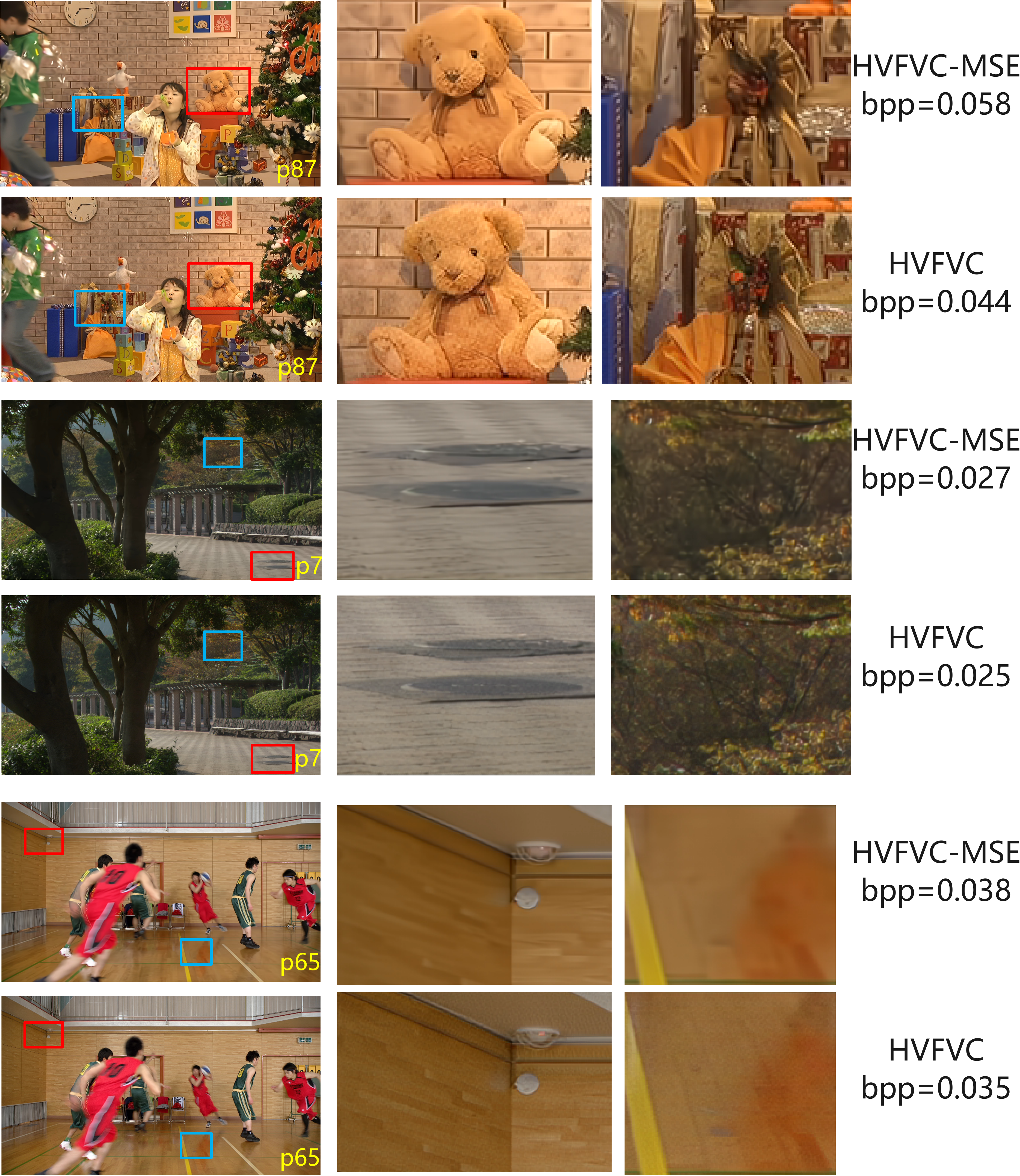}
	\caption{Comparison between reconstructed frames with HVFVC and HVFVC-MSE.}
	\label{fig:loss_cmp_sup}
\end{figure}

In this section, we compared the results of optimizing the architecture using MSE (HVFVC-MSE) and the mentioned perceptual loss (HVFVC). The visual results are shown in Fig. ~\ref{fig:loss_cmp_sup}. As can be seen in the figure, our perceptual loss can produce more detailed texture, and the reconstructed frames are more visual friendly.

\begin{figure*}[h]
	\centering
	\includegraphics[width=0.9\textwidth]{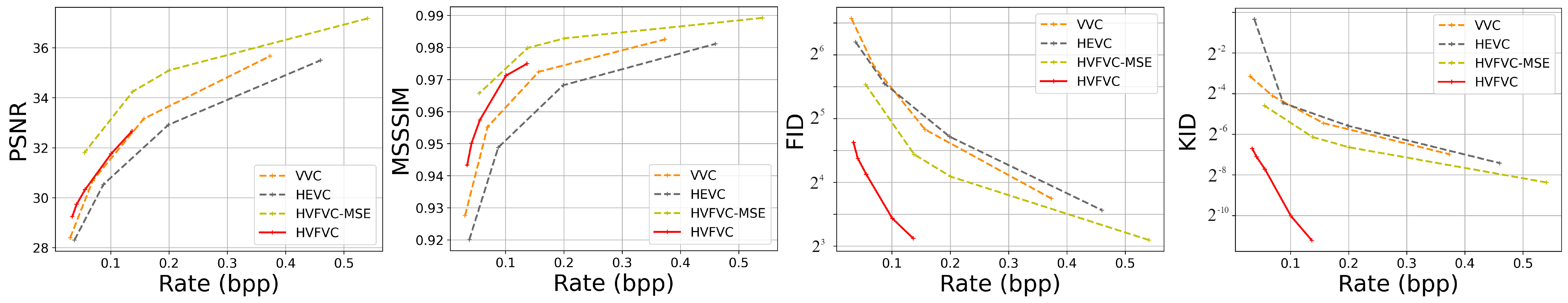}
	\caption{Statistic results of different methods.}
	\label{fig:rd_sup}
\end{figure*}

\begin{figure}[ht]
	\includegraphics[width=0.5\textwidth]{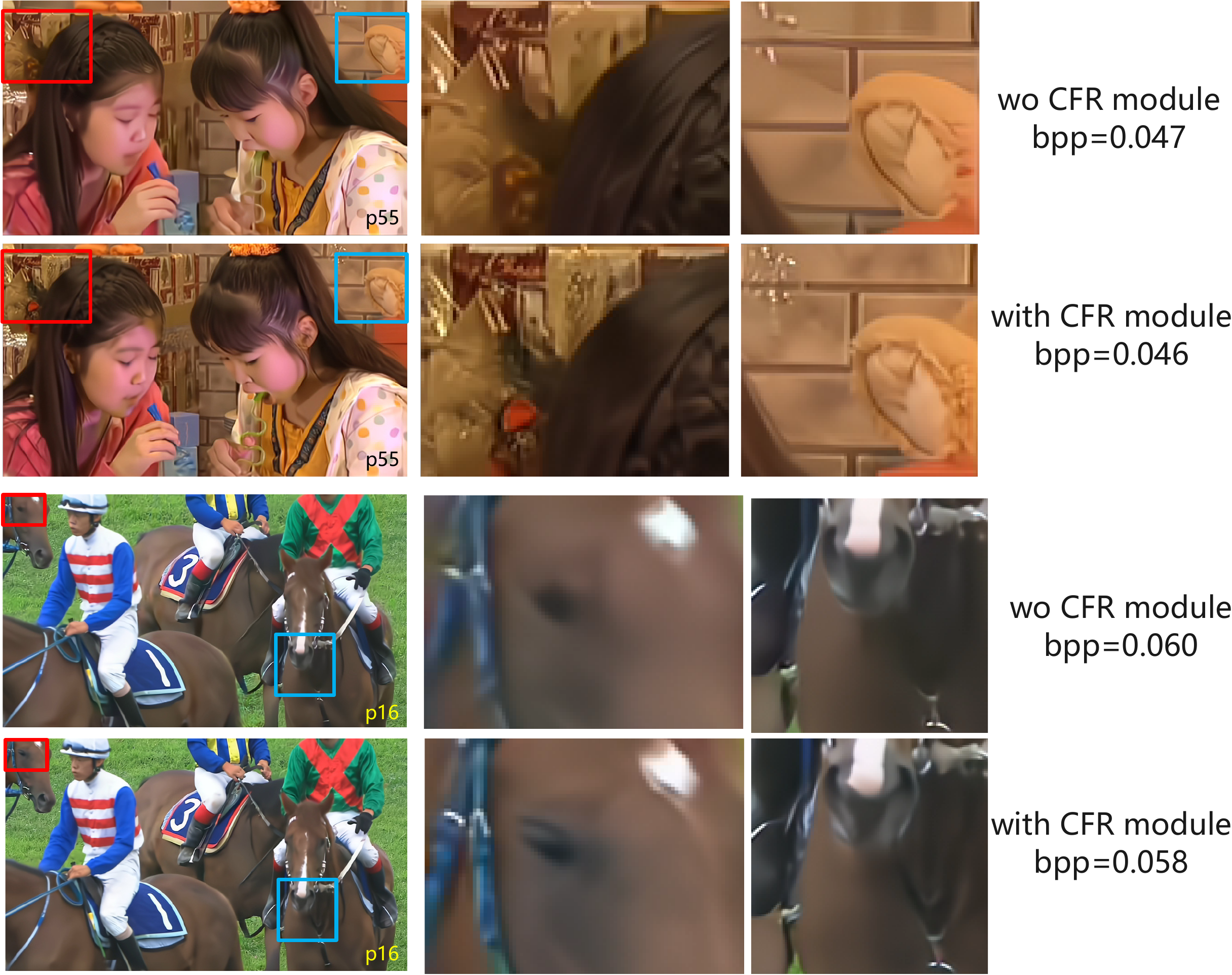}
	\caption{Additional ablation on the CFR module.}
	\label{fig:rcia_ablation_sup}
\end{figure}

\begin{figure}[t]
	\includegraphics[width=0.3\textwidth]{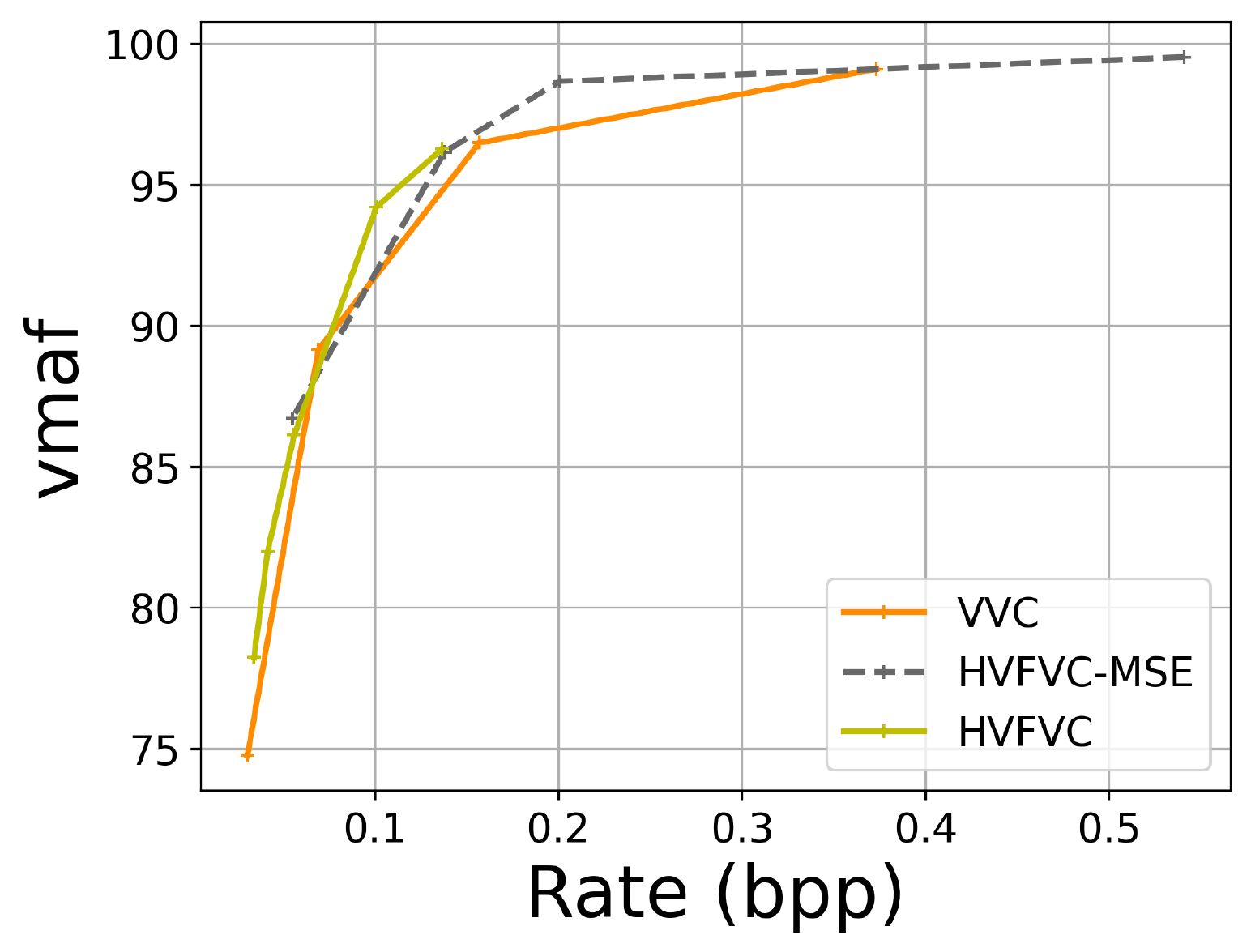}
	\caption{Comparison of VMAF between different methods.}
	\label{fig:VMAF}
\end{figure}

Besides, we also provide additional quantitative metrics to show the effectiveness of our method, including PSNR, MSSSIM, FID, and KID, the results of which can be found in Fig. ~\ref{fig:rd_sup}.  Compared to HEVC and VVC, our MSE-optimized model (HVFVC-MSE) outperforms both codecs across all metrics significantly, thus proving the superiority of our architecture. And the perceptual optimized model (HVFVC) exhibits superior performance over HEVC and VVC in terms of metrics including MSSSIM, FID and KID, indicating its effectiveness in achieving high visual quality. Moreover, HVFVC also achieves comparable PSNR performance to VVC while surpassing HEVC in this regard. Additionally, comparison of video quality metric VMAF (Video Multi-method Assessment Fusion) is also shown in Fig.~\ref{fig:VMAF} for reference.

\subsection{Complexity of HVFVC}
Our HVFVC has 1.39M  MACs(multiply-accumulate operations)  per pixel at the encoding side and 0.67M MACs per pixel at the decoding side. Table. ~\ref{time} shows the encoding and decoding times of different methods for a 1080p frame. The running platform of HVFVC is Intel(R) Xeon(R) Gold 6278C CPU and NVIDIA V100 GPU. Compared to HEVC and VVC, HVFVC has much shorter encoding time, but its decoding time is longer. Compared to other learned video compression methods, HVFVC and TCMVC have similar encoding and decoding times, which are both faster than DCVC.

\subsection{Effectiveness of the CFR module}
Addtionally, we also present more ablative experiment results on the confidence based feature reconstruction (CFR) module to further investigate its functionality and impact. The results are shown in Fig. ~\ref{fig:rcia_ablation_sup}. It is evident that with the help of the CFR module, the generated frames are more realistic, with richer textures.

\begin{table}[ht]
	\caption{Average encoding/decoding time for a 1080P frame (in seconds)}
	\begin{tabular}{c|c|c}
		\hline
		Method & Enc Time & Dec Time \\ \hline
		HEVC   & 26.47 s    & 0.14 s    \\ \hline
		VVC    & 661.90 s  & 0.22 s    \\ \hline
		DCVC   & 12.26 s   & 35.59 s    \\ \hline
		TCMVC  & 0.88 s    & 0.47 s    \\ \hline
		HVFVC  & 0.84 s    & 0.50 s    \\ \hline
	\end{tabular}
	\label{time}
\end{table}

\end{document}